\documentclass[12pt]{article}
\usepackage{amssymb}

\newcommand{\be}{\begin{equation}}
\newcommand{\ee}{\end{equation}}
\def\bea{\begin{eqnarray}}
\def\eea{\end{eqnarray}}

\def\mR{\mathbb{R}}

\def \barl {\bar{\Lambda}}
\def \brho {\bar{\rho}}
\def \trho {\tilde{\rho}}
\def \bphi {\bar{\phi}}
\def \tGamma {\tilde{\Gamma}}
\def \tgamma {\tilde{\gamma}}
\def \bGamma {\bar{\Gamma}}

\newcommand{\bn}{\begin{eqnarray}}
\newcommand{\en}{\end{eqnarray}}

\newcommand{\p}{\partial}

\newcommand{\nn}{\nonumber}

\newcommand{\no}{\noindent}

\newcommand{\s}{\,\,\,\,}
\def\bea{\begin{eqnarray}}
\def\eea{\end{eqnarray}}

\newcommand{\beq}{\begin{eqnarray}}
\newcommand{\eeq}{\end{eqnarray}}

\begin{document}

\title{\textbf{Higher order self-dual models for spin-3 particles in
$D=2+1$}}
\author{D. Dalmazi$^{1}$\footnote{dalmazi@feg.unesp.br}, A. L. R. dos Santos$^{2}$\footnote{alessandroribeiros@yahoo.com.br}, R. R. Lino dos Santos$^{1}$\footnote{rafa.robb@gmail.com} \\
\textit{{1- UNESP - Campus de Guaratinguet\'a - DFQ} }\\
\textit{{CEP 12516-410 - Guaratinguet\'a - SP - Brazil.} }\\
\textit{{2- Instituto Tecnol\'ogico de Aeron\'autica, DCTA} }\\
\textit{{CEP 12228-900, S\~ao Jos\'e dos Campos - SP - Brazil} }\\}
\date{\today}
\maketitle

\begin{abstract}

In $D=2+1$ dimensions, elementary particles  of a given helicity can be described by local Lagrangians (parity
singlets). By means of a ``soldering'' procedure two opposite helicities can be joined together  and give rise
to massive spin-$s$ particles carrying both helicities $\pm s$ (parity doublets), such Lagrangians can also be
used in $D=3+1$ to describe massive spin-$s$ particles. From this point of view the parity singlets (self-dual
models) in $D=2+1$  are the building blocks of real massive elementary particles in $D=3+1$. In the three cases
$s=1,\, 3/2,\, 2$ there are $2s$ self-dual models of order $1,2, \cdots, 2s$ in derivatives. In the spin-3 case
the 5th order model is missing in the literature. Here we deduce a 5th order spin-3 self-dual model and fill up
this gap. It is shown to be ghost free by means of a master action which relates it with the top model of 6th
order. We believe that our approach can be generalized to arbitrary integer spin-$s$ in order to obtain the
models of order $2s$ and $2s-1$. We also comment on the difficulties in relating the 5th order model with their
lower order duals.

\end{abstract}

\newpage

\section{ Introduction}

Although we have not yet seen higher spin ($s \ge 3/2$) elementary particles in nature, massive  particles of arbitrarily
high spin are predicted by string theory. It is tempting to connect the non detection of such particles to the
theoretical difficulties we have in formulating a self-consistent theory where such particles interact with
themselves or with other fields like the gravitational field. As we increase the spin we need higher rank
tensors. However, only $2s +1$ degrees of freedom should survive for a massive spin-s
particle in $D=3+1$ dimensions. Consequently, several spurious fields must be consistently eliminated which
becomes cumbersome specially when interactions are present. The tensor fields must obey the so called
Fierz-Pauli constraints \cite{fp}.

Even the Lagrangians for free  particles must be fine tuned for higher spins in order to produce the correct
constraints \cite{sh}. We believe that those Lagrangians could be systematically obtained from a ``soldering''
procedure of opposite helicities in $D=2+1$ space-time, see more on soldering in
\cite{stone,bk,iw,review,gs,dm1,dm2}. In $D=2+1$ dimensions, differently from the real world, it is possible to
write down local Lagrangians for elementary particles of given helicity. Helicity eigenstates are described by
parity singlets, the so called self-dual (SD) models in $D=2+1$. For instance, two Maxwell-Chern-Simons (MCS)
theories (spin-1) of helicities $+1$ and $-1$, suggested in \cite{djt}, can be soldered into the spin-1
Maxwell-Proca model whose action has the same form in arbitrary dimensions. Likewise, two spin-2 self-dual
models of helicities $+2$ and $-2$ of second order\footnote{Throughout this work $n$-th order model stands for a
Lagrangian whose maximal number of derivatives is $n$.} in derivatives, suggested in \cite{desermc}, can be
soldered into the massive spin-2 Fierz-Pauli (FP) theory. The fine tuned mass term of the FP theory is
automatically generated. It works also for higher derivative models. In particular, the  relative $-3/8$ factor
between $R_{\mu\nu}^2$ and $R^2$ terms in the linearized version of the ``New Massive Gravity'' of \cite{bht}
can also be automatically produced by the soldering of two linearized topologically massive gravities (LTMG) of
opposite helicities $+2$ and $-2$, see \cite{dm1}. The case of spin-3/2 has also been recently achieved
\cite{mls}. The spin-3 case is still under investigation.

From a constructive point of view, we believe that the self-dual models in $D=2+1$ are the building blocks for
massive particles in the real world. Therefore, it is certainly interesting to learn how to build them
systematically for arbitrarily higher spins, specially their higher order versions. As we increase the number of
derivatives, the number of FP constraints which are {\it identically} satisfied as opposed to {\it dynamically}
satisfied also increases which makes easier in principle the introduction of interactions without destroying the
constraints.

The higher derivative self-dual models can be obtained from their first order counterparts order by
order\footnote{The notation ${\cal L}_{SDn}^{(s)}$ stands for the Lagrangian density of the spin-s
self-dual model of order $n$ in derivatives.} (${\cal
L}_{SDj}^{(s)} \to {\cal L}_{SD(j+1)}^{(s)}$) in, at least, two different ways. Either by means of a master
action approach \cite{dj} (spin-1), see \cite{dual,sd4} for the spin-2 case, or via a Noether gauge embedding (NGE) procedure where the amount of local symmetries increases along with the number of derivatives, see \cite{clovis} (spin-1), \cite{msl} (spin-3/2) and \cite{sd4} (spin-2). Those three cases are consistent with a ``$2s$'' rule for the highest possible order in derivatives of a ghost-free  self-dual model for spin-$s$ particles. The case of spin-3 investigated here is quite challenging.
Although the sixth order model is known \cite{bhth}, starting with the first order model of \cite{ak1} we have obtained in \cite{nge} and \cite{dmmaster}  the second \cite{ak2}, the third \cite{deserdam} and a fourth order spin-3 self-dual model via the NGE and the master action procedures respectively. Although they are all ghost-free, we have not been able to go beyond the fourth order and reach the sixth order model. The NGE procedure requires more symmetry in the higher order term than in the rest of the Lagrangian which is not the case in the ${\cal L}_{SD4}^{(3)}$ model found in \cite{nge}. Regarding the master action, the highest order term in the Lagrangian can not have any particle content which is not the case either of ${\cal L}_{SD4}^{(3)}$ as we have shown in \cite{dmmaster}, see also section 4 in the present work. Here we tackle that problem by going downward from the sixth order theory \cite{bhth} and finding a ghost-free fifth order model (section 2). In section 3 we connect it with the sixth order model via master action. In section 4 we investigate a class of fourth order Lagrangians in search for a possible fourth order self-dual  model different from ${\cal L}_{SD4}^{(3)}$ of \cite{dmmaster} which would allow us to go further downward from ${\cal L}^{(3)}_{SD5}$. In section 5 we have our conclusions and perspectives.

\section{The fifth order self-dual model ${\cal L}_{SD5}^{(3)}$ }

We start this section recalling the construction of the sixth order spin-3 self-dual model $SD^{(3)}_6$ of
\cite{bhth}. We follow a route  slightly different from \cite{bhth} which we think it could be more easily
generalized to arbitrary integer spins. First, for the spin-1 and spin-2 cases the highest self-dual models of order
$2s$ are given respectively by the Maxwell-Chern-Simons \cite{djt} and the linearized higher derivative
topologically massive gravity of \cite{sd4,andringa}. They can be written in a compact way with the help of dual
fields $h^*$. Namely\footnote{ We use $\eta_{\mu\nu} = (-,+,+)$, $(\alpha\beta)=\alpha\beta +
\beta\alpha $ and $(\alpha\beta\gamma)=\alpha\beta\gamma + \beta\gamma\alpha + \gamma\alpha\beta  $.},

\bea {\cal L}_{SD2}^{(1)}&=&\frac{m}{2}h_{\mu}E^{\mu\nu}h_{\nu}^* -
\frac{m^2}2 h^{\mu}h_{\mu}^* \quad,\nn\\
{\cal
L}_{SD4}^{(2)}&=&\frac{m}{2}h_{\mu}^{\s\nu}E^{\mu\alpha}h^*_{\alpha\nu}-\frac{m^2}{2}h^{\mu\nu}h_{\mu\nu}^*\quad,\label{sdn}\eea

\no where the dual fields are given by

\bea h_{\mu}^* &=& E_{\mu\nu}h^{\nu}/m \quad ,  \label{dualf1} \\
h_{\mu\nu}^*
&=&\,\left(E_{\mu}^{\s\alpha}\Box\theta_{\nu}^{\s\beta}+E_{\nu}^{\s\alpha}\Box\theta_{\mu}^{\s\beta}\right)
h_{\alpha\beta}/(2m^3) \quad . \label{dualf2}\eea

\no The transverse operators

\be E^{\rho\delta} \equiv \epsilon^{\rho\delta\sigma}\p_{\sigma} \quad ; \quad  \Box \theta_{\rho\sigma} \equiv
\Box \eta_{\rho\sigma} - \p_{\rho} \p_{\sigma}\quad, \label{etheta} \ee

\no are such that

\be  E^{\mu\nu} E ^{\alpha\beta} = \Box \left( \theta^{\mu\beta}\theta^{\nu\alpha} -
\theta^{\mu\alpha}\theta^{\nu\beta} \right) \quad . \label{id1} \ee

\no The dual fields identically satisfy the respective Fierz-Pauli (FP) constraints:

\be \p^{\mu}h_{\mu}^* = 0 \quad ; \quad \p^{\mu}h_{\mu\nu}^* =0 \quad ; \quad \eta^{\mu\nu}h_{\mu\nu}^* =0\quad.
\label{fpc12} \ee

\no The equations of motion are given respectively by

\bea  E_{\mu}^{\s\alpha}h^*_{\alpha} &=& m\, h^*_{\mu}\quad, \label{eoms1}\\
 E_{(\mu}^{\s\alpha}h^*_{\alpha\nu )} &=& 2\, m \, h^*_{\mu\nu}\quad. \label{eoms2} \eea

\no By applying $E^{\beta\mu}$ on the equations of motion, using (\ref{id1}),(\ref{fpc12}), equations of
motion recursively and symmetrizing the result (only in the spin-2 case of course) we derive the Klein-Gordon
(KG) equations :

\be (\Box - m^2)h_{\beta}^* = 0 \quad ; \quad (\Box - m^2)h_{\beta\nu}^* =0 \quad . \label{kg12} \ee

\no The Pauli-Lubanski equations (\ref{eoms1}),(\ref{eoms2}) single out one helicity eigenstate. They form
altogether with (\ref{fpc12}) and (\ref{kg12}) all the required equations for helicity-$s$ particles ($s=1,2$) in
$D=2+1$ dimensions (parity singlets) represented by the dual field $h^*$. The natural generalization of
(\ref{sdn}) for spin-3 would be a sixth order self-dual model for a totally symmetric rank-3 tensor,

\be {\cal
L}_{SD6}^{(3)}=\frac{m}{2}h_{\mu}^{\s\nu\rho}E^{\mu\alpha}h^*_{\alpha\nu\rho}-\frac{m^2}{2}h^{\mu\nu\rho}h_{\mu\nu\rho}^* \quad.\label{sd6}\ee

\no The totally symmetric dual field $h_{\mu\nu\rho}^*$ is obtained by applying some fifth order differential
operator on $h_{\mu\nu\rho}$ such that the required spin-3 Fierz-Pauli constraints are identically satisfied:

\be \p^{\mu}h_{\mu\nu\rho}^* =0 \quad ; \quad \eta^{\mu\nu}h_{\mu\nu\rho}^* =0 \quad. \label{fpc3} \ee

\no If we apply $E_{\beta}^{\s\mu}$ on the equations of motion:

\be E_{(\mu}^{\s\alpha}h^*_{\alpha\nu\rho )} = 3\, m \, h^*_{\mu\nu\rho} \quad, \label{eoms3} \ee

\no and use (\ref{id1}), (\ref{fpc3}) and (\ref{eoms3}) recursively we obtain the expected KG equations $(\Box - m^2)h^*_{\beta\nu\rho}=0$.

 The invariance of the third order Lagrangian $h^{\mu\nu}h_{\mu\nu}^*$  under linearized reparametrizations and Weyl transformations $\delta h_{\mu\nu} = \p_{(\mu}\Lambda_{\nu )} + \eta_{\mu\nu} \psi $ is enough to guarantee that the spin-2 FP constraints in (\ref{fpc12}) hold identically.
Likewise, we require invariance of the fifth order Lagrangian
 $h^{\mu\nu\rho}h_{\mu\nu\rho}^*$ under:

\be \delta h_{\mu\nu\rho} = \p_{(\mu}\Lambda_{\nu\rho )} + \eta_{(\mu\nu} \psi_{\rho )} \quad.\label{gt3}\ee

\no The first symmetry imply that the six indices of the two $h_{\mu\nu\rho}$ fields present in
$h^{\mu\nu\rho}h_{\mu\nu\rho}^*$ be contracted with indices of transverse operators $E^{\mu\nu}$ or $\Box
\theta^{\mu\nu}$. There are only two possibilities at fifth order:

\be {\cal L}_{AB} = h^{\mu\nu\rho}\Box^2 E_{\mu}^{\s\alpha} \left( A \, \theta_{\nu}^{\beta}
\theta_{\rho}^{\gamma} + B \, \theta_{\nu\rho}\theta^{\beta\gamma} \right)h_{\alpha\beta\gamma} \quad .
\label{lab} \ee

\no The vector Weyl invariance then fixes $B=-A/4$.  Therefore, the sixth order spin-3 self-dual model, in
agreement with \cite{bhth}, is given by (\ref{sd6}) where

\be h_{\mu\nu\rho}^* = \frac{\Box^2}{3m^{5}}  E_{(\mu}^{\s\alpha} \left( \theta_{\nu}^{\beta} \theta_{\rho )}^{\gamma} -  \frac
14 \theta_{\nu\rho )}\theta^{\beta\gamma} \right) h_{\alpha\beta\gamma}    \quad . \label{dualf3} \ee

\no Alternatively, one can
start with a rather general Ansatz for a fifth order Lagrangian including all possible contractions (five
terms), the symmetry (\ref{gt3}) will  finally lead us to the same answer.

Both sixth and fifth order terms in ${\cal L}_{SD6}^{(3)}$ are invariant under the same set of gauge
transformations (\ref{gt3}), this is the typical situation in the highest (2s) order self-dual models, see
\cite{sd4}. The high degree of symmetry avoids the presence of ghosts which commonly appear in higher derivative
theories. The absence of ghosts in ${\cal L}_{SD6}^{(3)}$ has been explicitly proven in \cite{bhth} in the gauge
$\p_{j}h_{j\mu\nu} = 0$. They have also shown that the fifth order term $h^{\mu\nu\rho}h_{\mu\nu\rho}^*$ has by
itself no particle content  very much like the third order linearized gravitational Chern-Simons term in the
spin-2 case, see \cite{djt}. Since the later one can be combined with the linearized Einstein-Hilbert action
(another empty theory) in order to produce a meaningful spin-2 self-dual model (linearized topologically massive
gravity), one naturally wonders whether we could combine $h^{\mu\nu\rho}h_{\mu\nu\rho}^*$ with some fourth order
term and end up with a ghost free fifth order self-dual model.

As we have already mentioned, there is a systematic procedure to go from the $n$-th to the $(n+1)$-th order
self-dual model by either using the embedding of gauge symmetries or a master action \cite{dj} approach.
However, to the best we know there is no systematic way to go downward in derivatives. In the spin-2 and spin-1
cases both self-dual models of order $2s-1$, i.e.,  ${\cal L}_{SD3}^{(2)}$ and ${\cal L}_{SD1}^{(1)}$, are built
up from two terms of zero particle content. This is a key ingredient in the master action approach. So we must
seek for a highly symmetric fourth order term. Starting with a general Ansatz:

\bea {\cal L}^{(4)} &=& a\, h_{\mu\nu\alpha}\Box^2 h^{\mu\nu\alpha} + b \, h_{\mu}\Box^2 h^{\mu} + c \, h_{\mu\nu\alpha}\Box \p^{\mu}\p_{\rho}h^{\rho\nu\alpha} + d \, h_{\mu}\Box \p^{\mu}\p_{\nu}h^{\nu} \nn\\
&&+ e \, h_{\mu\nu\alpha} \p^{\mu}\p^{\nu}\Box h^{\alpha} + f \,
\p^{\mu}\p^{\nu}h_{\mu\nu\alpha}\p_{\gamma}\p_{\rho}h^{\gamma\rho\alpha} + g \,  \p^{\mu}\p^{\nu}\p^{\alpha}
h_{\mu\nu\alpha} \p_{\rho}h^{\rho} \quad.\label{ansatz4} \eea

\no Let us first define a subclass of models invariant under traceless reparameterizations\footnote{There is one case where we have invariance under arbitrary reparametrizations  $\delta{h}_{\mu\nu\alpha}=\partial_{(\mu}\lambda_{\nu\alpha)}$, however it coincides with the fourth order term appearing in \cite{nge} which contains a ghost.}:

\be \delta h_{\mu\nu\rho} = \p_{(\mu}\bar{\Lambda}_{\nu\rho )} \quad ; \quad \eta^{\mu\nu}
\bar{\Lambda}_{\mu\nu} = 0 \quad . \label{lbar} \ee

 \no Such symmetry can be implemented if

 \be c=-3\, a \quad ; \quad d = \frac 14 (9\, a + 5 \, b) \quad ; \quad e= -2 \, b \quad ; \quad f=g=3\, a + b \quad.\label{ctes} \ee

 \no Accordingly, (\ref{ansatz4}) becomes

\be {\cal L}^{(4)}_{(a,b)} = a \left(\mR_{\mu\nu\rho}\mR^{\mu\nu\alpha} - \frac 34 \mR_{\mu}\mR^{\mu}\right) + \frac b4
\mR_{\mu}\mR^{\mu} \quad,\label{lab} \ee

\no where the spin-3 Ricci-like curvature and its vector contraction have been introduced in \cite{deserdam},
namely,

\bea \mR_{\mu\nu\rho} &=& \Box h_{\mu\nu\rho} - \p^{\alpha}\left(\p_{\mu} h_{\alpha\nu\rho} + \p_{\nu}h_{\alpha\mu\rho} + \p_{\rho} h_{\alpha\mu\nu} \right) + \p_{\mu}\p_{\nu} h_{\rho} + \p_{\nu}\p_{\rho}h_{\nu} + \p_{\rho}\p_{\mu} h_{\nu} \quad,\nn \\
\mR_{\rho} &=& \eta^{\mu\nu}\mR_{\mu\nu\rho} = 2 \left( \Box h_{\rho} - \p^{\alpha}\p^{\beta} h_{\alpha\beta\rho}
+ \frac 12 \p^{\alpha}\p_{\rho} h_{\alpha} \right) \quad. \label{ricci} \eea

\no Note that $\mR_{\mu\nu\rho}$, and consequently $\mR_{\mu}$,  is invariant under (\ref{lbar}). The traceless
reparametrization symmetry  (\ref{lbar}) plays in the massive spin-3 theory in $D=2+1$, and also in $D=3+1$
\cite{sh}, the same role of the linearized reparametrizations $\delta h_{\mu\nu} = \p_{(\mu}\Lambda_{\nu )} $ in
the spin-2 FP theory, i.e., it is the symmetry of the massless limit of the theory and is instrumental in deriving the FP constraints in the massive case.

Now we are ready to suggest an Ansatz for the fifth order spin-3 self-dual model. Usually, the lower derivative term of the $n$-th order self-dual model becomes the higher derivative term of the ($n-1$)-th model but with opposite sign. Thus, the second term of (\ref{sd6}), with opposite sign, now becomes the highest order term of ${\cal L}_{SD5}^{(3)}$. We add up the fourth order term (\ref{lab}) and require the final Lagrangian to fit into the form (\ref{sd6}) in terms of a fourth order dual field $\tau_{\mu\nu\rho}^*$:

\be {\cal L}_{SD5}^{(3)}\; =\; \frac{m^2}2 h^{\mu\nu\rho}h_{\mu\nu\rho}^* - {\cal L}^{(4)}_{(a,b)} \;=\; \frac m2
h_{\mu}^{\s\nu\rho}E^{\mu\alpha}\tau^*_{\alpha\nu\rho}- \frac{m^2}2 h^{\mu\nu\rho}\tau_{\mu\nu\rho}^* \quad.
\label{sd5} \ee

\no From (\ref{sd5}) and (\ref{dualf3}) we end up with a unique solution :

\be a= \frac 1{2\, m^2} \quad ; \quad b = -\frac 34 \, a = - \frac 3{8\, m^2} \quad . \label{tw}\ee

\no Explicitly, the fourth order dual field is given by

\bea \tau_{\mu\nu\rho}^* \!&=&\!\frac{1}{m^4} \Bigg[\Box^2 h_{\mu\nu\rho} - \frac 14 \eta_{(\mu\nu}\Box^2 h_{\rho )} - \Box \p_{(\mu}\p^{\beta}h_{\beta\nu\rho )} + \frac{\Box}4 \eta_{(\mu\nu}\p_{\gamma}\p_{\beta}h_{\rho )}^{\s\gamma\beta} + \frac{\Box}4 \p_{(\mu}\p_{\nu}h_{\rho )} \nn\\
&& + \frac{7}{16} \eta_{(\mu\nu}\p_{\rho )} \Box \p \cdot h + \frac 34 \p_{(\mu}\p_{\nu}\p^{\gamma}\p^{\beta}h_{\gamma\beta\rho )} - \frac 98 \p_{\mu}\p_{\nu}\p_{\rho}(\p \cdot h) - \frac 38 \eta_{(\mu\nu}\p_{\rho)}\p^{\gamma}\p^{\beta}\p^{\delta}h_{\gamma\beta\delta} \Bigg]\;. \nn\\
\label{tau}\eea

\no It turns out that the fourth order term in (\ref{sd5}) is exactly the same one appearing  in the fourth
order description of massive spin-3 particles (parity doublet) of \cite{dmh}.
It can be written as the product
of a spin-3 second order Einstein-like and a Schouten-like tensor very much like the fourth order term (K-term)
of the ``New Massive Gravity'' (NMG) theory \cite{bht}. After integrating by parts we can write:

\bea
{\cal L}_{SG} \equiv \frac{m^2}{2} h^{\mu\nu\rho}\tau_{\mu\nu\rho}^{\ast} = \frac{1}{2m^{2}}\mathbb{S}_{\mu\nu\alpha}(h)\mathbb{G}^{\mu\nu\alpha}(h)=\frac{1}{2m^2}\mathbb{R}_{\mu\nu\lambda}\mathbb{R}^{\mu\nu\lambda}-\frac{15}{32\, m^{2}}\mathbb {R}_{\mu}\mathbb{R}^{\mu}\quad,\eea

\no where, see \cite{deserdam},

\be {\mathbb G}_{\mu\nu\lambda}\equiv {\mathbb R}_{\mu\nu\lambda}-\frac{1}{2}\eta_{(\mu\nu}{\mathbb R}_{\lambda
)} \quad ; \quad {\mathbb S}_{\mu\nu\lambda} = {\mathbb G}_{\mu\nu\lambda}-\frac 14 \eta_{(\mu\nu}{\mathbb
G}_{\lambda )}={\mathbb R}_{\mu\nu\lambda} -\frac{1}{8}\eta_{(\mu\nu}{\mathbb R}_{\lambda )}\quad.\label{GS}\ee

\no The definition of the Einstein-like tensor is consistent with the traceless reparametrization in $D=2+1$ in
the sense that ${\mathbb G}_{\mu\nu\lambda}(h)=0$ implies  a pure gauge solution $h_{\mu\nu\rho} =
\p_{(\mu}\bar\Lambda_{\nu\rho )}$ just like $G_{\mu\nu}(h)=0$ leads to $h_{\mu\nu} = \p_{(\mu}\Lambda_{\nu )}$ in the spin-2 case.
Besides traceless reparametrizations (\ref{lbar}), the fourth order term ${\cal L}_{SG}$ is the only possible
combination among all fourth order terms (\ref{ansatz4}) which is invariant also under transverse Weyl transformation (quite similar again to the K-term in NMG)

\be \delta h_{\mu\nu\rho} = \eta_{( \mu\nu} \psi_{\rho )}^T \quad ; \quad \p^{\mu}\psi_{\mu}^T = 0  \quad.
\label{transw} \ee

\no The fifth order term in (\ref{sd5}) can also be written in a more inspiring form, namely,

\be \frac{m^2}2 h^{\mu\nu\rho}h_{\mu\nu\rho}^* =
\frac{1}{4m^{3}}\mathbb{S}_{\mu\nu\alpha}(h)\mathbb{G}^{\mu\nu\alpha}(Eh)\quad, \label{seg} \ee

\no where $(Eh)_{\mu\nu\alpha}=(2/3)E_{(\mu\beta}h^{\beta}_{\,\,\nu\alpha)}$. Notice that the fifth order
term in ${\cal L}_{SD5}^{(3)}$ is invariant under (\ref{lbar}) and full Weyl transformations, see (\ref{gt3}).
It is a common feature of lower order (below $2s$)  self-dual models that the highest derivative term has more
symmetries than its lower derivative partner. Next we show  via master action that the particle content of ${\cal L}_{SD5}^{(3)}$ is the same one of ${\cal L}_{SD6}^{(3)}$, i.e., helicity $+3$ or $-3$ particles, depending on the sign of the fifth order term, without ghosts.

 The generalization of the 6th order model (\ref{sd6}) for arbitrary spin-$s$ goes in the following way. We first replace $h_{\mu\nu\rho}$ by a rank-$s$ totally symmetric field $h_{\alpha_1 \cdots \alpha_s}$. The dual field $h_{\alpha_1 \cdots \alpha_s}^*$ is built up out of a differential operator of order $2s-1$ applied on $h_{\alpha_1 \cdots \alpha_s}$  such that the rank-s generalization of (\ref{gt3}) becomes a symmetry of  
$h^{\alpha_1 \cdots \alpha_s} h_{\alpha_1 \cdots \alpha_s}^* $. As we increase the spin we have more terms which contribute, however the rank-s  version of (\ref{gt3}) is enough \cite{ddos} to uniquely fix the dual field $h_{\alpha_1 \cdots \alpha_s}^*$. The factor $3\, m$ will be replaced by $s\, m$ on the right hand side of (\ref{eoms3}). We believe that the lower order $2s-1$ self-dual model can be defined for arbitrary spin-s following the same route of the 5th order spin-3 model. Namely, we define a term of order $2s-2$ by requiring traceless reparametrizations and then impose that its  symmetrized curl becomes the already known $2s-1$ term $h^{\alpha_1 \cdots \alpha_s} h_{\alpha_1 \cdots \alpha_s}^* $. This is in progress \cite{ddos}. The increasing number of derivatives does not lead to ghosts due to the increasing number of local symmetries. The proof of absence of ghosts is still cumbersome due to the higher derivatives.

\section{Master action: $S^{(3)}_{SD5}\rightarrow{S}^{(3)}_{SD6}$}

In this section we use the master action technique to connect ${\cal L}_{SD5}^{(3)}$ with ${\cal L}_{SD6}^{(3)}$. We start with the  $S_{SD5}$ action and add a mixing term between the old field and the new dual field.
The mixing term is the fifth order term of $S_{SD5}$.

The $SD5$ action is given by
\bea S^{(3)}_{SD5}[h]=\int{d^{3}x}\Big[-\frac{1}{2m^{2}}\mathbb{S}_{\mu\nu\alpha}(h)
\mathbb{G}^{\mu\nu\alpha}(h)+\frac{1}{4m^{3}}\mathbb{S}_{\mu\nu\alpha}(h)\mathbb{G}^{\mu\nu\alpha}(Eh)\Big]\quad.\label{ssd5}\eea

It has been shown in \cite{bhth} that the fifth order term has no particle content. The fourth order term ${\cal
L}_{SG}$ has also an empty spectrum as we have shown in \cite{dmh} via a duality transformation to a rank-2
theory, we will say more about it later on. Following the master action procedure, we use the fifth order term
as mixing to construct the following master action:
\bea
S_{M}[h,f]=S^{(3)}_{SD5}[h]-\frac{1}{4m^{3}}\int{d^{3}x}\;\mathbb{S}_{\mu\nu\alpha}(h-f)\mathbb{G}^{\mu\nu\alpha}\big(E(h-f)\big) \quad,\label{sm}
\eea
where $f_{\mu\nu\alpha}$ is the new field introduced through the mixing term. In order to find a dual map between
correlation functions of the dual models, we add a source term $j^{\mu\nu\alpha}$ coupled to a totally symmetric
dual field $\tau^{\ast}_{\mu\nu\alpha}$ and define the following generating functional

\be Z_{M}[j]=\int\mathcal{D}h\mathcal{D}f\,\mbox{exp}\,i\Big(S_{M}[h,f]+\int{d^{3}x}\,
\tau^{\ast}_{\mu\nu\alpha}j^{\mu\nu\alpha}\Big) \quad.\ee
The dual field is given by
$\tau^{\ast}_{\mu\nu\alpha}=(1/m^{4})\mathbb{G}_{\mu\nu\alpha}(\mathbb{S}(h))$, see explicitly in (\ref{tau}).
It guarantees the gauge symmetry of the master action
under $\delta{h}_{\mu\nu\alpha}=\partial_{(\mu}\bar\Lambda_{\nu\alpha)}+\eta_{(\mu\nu}\psi^{T}_{\alpha)}$.

After making the shift $f\rightarrow{f}+h$ in (\ref{sm}) we recover the fifth order self-dual model plus a
decoupled fifth order term depending only on $f_{\mu\nu\rho}$. Since the fifth order term has no particle
content, this guarantees that the particle content of $S_{M}[h,f]$ is the same of $ S^{(3)}_{SD5}[h]$.  On
the other hand, the action can be written as \footnote{We have used the following properties in the source term: $\mathbb{G}_{\mu\nu\alpha}(\mathbb{S}(h))j^{\mu\nu\alpha}=\mathbb{G}_{\mu\nu\alpha}(h)\mathbb{S}^{\mu\nu\alpha}(j)=\mathbb{S}_{\mu\nu\alpha}(h)\mathbb{G}^{\mu\nu\alpha}(j)$.}

\bea S_{M}\!&=&\! \int{d^{3}x}\Big[-\frac{1}{4m^{3}}\mathbb{S}_{\mu\nu\alpha}(f)\mathbb{G}^{\mu\nu\alpha}(Ef)-\frac{1}{2m^{2}}\mathbb{S}_{\mu\nu\alpha}(h)\mathbb{G}^{\mu\nu\alpha}(h)\nn\\
&&\qquad\;\quad+\frac{1}{2m^{3}}\mathbb{S}_{\mu\nu\alpha}(h)\mathbb{G}^{\mu\nu\alpha}(Ef)+\frac{1}{m^{4}}\mathbb{S}_{\mu\nu\alpha}(h)\mathbb{G}^{\mu\nu\alpha}(j)\Big] \quad,\eea

\no which is equivalent to

\bea S_{M} \!&=&\! \int{d^{3}x}\Bigg[-\frac{1}{4m^{3}}\mathbb{S}_{\mu\nu\alpha}(f)\mathbb{G}^{\mu\nu\alpha}(Ef)+\frac{1}{8m^{4}}\mathbb{S}_{\mu\nu\alpha}(Ef)\mathbb{G}^{\mu\nu\alpha}(Ef)+\frac{1}{4m^{5}}\mathbb{G}_{\mu\nu\alpha}(Ef)\mathbb{S}^{\mu\nu\alpha}(j)\nn\\
&&\qquad\;\quad -\frac{1}{2m^{2}}\mathbb{S}_{\mu\nu\alpha}\Big(h- \frac{Ef}{2m}-\frac{j}{m^{2}} \Big)\mathbb{G}^{\mu\nu\alpha}\Big(h-
\frac{Ef}{2m}-\frac{j}{m^{2}} \Big) \Bigg] \quad.\label{sm3} \eea

\no After making the shift

\bea h_{\mu\nu\alpha}\rightarrow
h_{\mu\nu\alpha}+\frac{1}{2m}(Ef)_{\mu\nu\alpha}+\frac{{j}_{\mu\nu\alpha}}{m^{2}} \quad , \eea

\no the last term of (\ref{sm3}) decouples. Since such term has no particle content it can be trivially Gaussian
integrated. Thus, we finally obtain the sixth order self-dual model given by \footnote{We have used the
property:
$\mathbb{S}_{\mu\nu\alpha}(Ef)\mathbb{G}^{\mu\nu\alpha}(Ef)=\mathbb{S}_{\mu\nu\alpha}(f)\mathbb{G}^{\mu\nu\alpha}(E^{2}f)$.}
\bea S^{(3)}_{SD6}[f]\!&=&\!\int{d^{3}x}\Big[-\frac{1}{4m^{3}}\mathbb{S}_{\mu\nu\alpha}(f)\mathbb{G}^{\mu\nu\alpha}(Ef)+\frac{1}{8m^{4}}\mathbb{S}_{\mu\nu\alpha}(f)\mathbb{G}^{\mu\nu\alpha}(E^{2}f)+f^{\ast}_{\mu\nu\alpha}j^{\mu\nu\alpha}+\mathcal{O}(j^{2})\Big]\;,\nn\\
\label{ssd6}\eea
where $f^{\ast}_{\mu\nu\alpha}=(1/2m^{5})\mathbb{G}_{\mu\nu\alpha}(\mathbb{S}(Ef))$ is the dual field $h^{\ast}_{\mu\nu\alpha}$ given in (\ref{dualf3}) after the replacement $h_{\mu\nu\alpha}\rightarrow{f}_{\mu\nu\alpha}$.
The equivalence between the $S^{(3)}_{SD5}$ and $S^{(3)}_{SD6}$ actions is guaranteed by the master action $S_{M}$. So
the action (\ref{ssd6}) describes helicity +3 (or $-3$) eigenmodes depending on the sign of the 5th order term. The correlation functions in both theories are related
by

\be \langle \tau^*_{\mu_{1}\nu_{1}\alpha_{1}}\ldots \tau^{\ast}_{\mu_{N}\nu_{N}\alpha_{N}} \rangle_{SD5}=\langle
{f}^{\ast}_{\mu_{1}\nu_{1}\alpha_{1}}\ldots{f}^{\ast}_{\mu_{N}\nu_{N}\alpha_{N}}\rangle_{SD6}+C.T.\quad,\ee

\no where $C.T.$ are contact terms which appear due to the quadratic terms on the sources in the master action.
The equations of motion of ${\cal L}_{SD5}^{(3)}(h)$ are mapped in the equations of motion of ${\cal
L}_{SD6}^{(3)}(f)$ via the substitution

\be \tau^{\ast}_{\mu\nu\alpha}\;\to\; {f}^{\ast}_{\mu\nu\alpha} \quad.\ee

\no The equivalence between the new fifth order spin-3 self-dual model given in (\ref{sd5}) or (\ref{ssd5}) and the sixth
order model of \cite{bhth} rises up the question whether there might be another master action allowing us to go
further downwards to reach a fourth order self-dual model and eventually fill up the gap in the chain of
self-dual models found in \cite{nge} from ${\cal L}_{SD1}^{(3)}$ to ${\cal L}_{SD4}^{(3)}$. In the next
section we investigate this issue by studying the particle content of the family of fourth order models
(\ref{lab}).

\section{ The particle content of ${\cal L}_{(a,b)}$}

A key ingredient in the master action approach is the fact that both the higher and the lower derivative terms
have no particle content. It turns out that the fourth order term present in the fourth order self-dual model of
\cite{dmmaster} contains two degrees of freedom and one of them
 is a ghost. This was an obstacle to go beyond the fourth order self-dual model in \cite{dmmaster}. The fourth order term of \cite{dmmaster} corresponds to ${\cal L}_{(a,b)}$ with $b=-a$ while the one we have used in (\ref{sd5})  corresponds to  $b=-3\, a/4$. The later  has no particle content. In order to clarify this issue and investigate any other possibility we examine the particle content of ${\cal L}_{(a,b)}$. As a by-product we offer an alternative proof of absence of content in the $b=-3\, a/4$ case.

Explicitly the ${\cal L}_{(a,b)}$ model (\ref{lab}) is given by

\bea {\cal L}_{(a,b)} &=& a \, h_{\mu\nu\rho} \Box^2 h^{\mu\nu\rho} + 3\, a \p^{\mu}h_{\mu\nu\rho} \Box \p_{\lambda}h^{\lambda\nu\rho} + b \, \left( h_{\mu} \Box^2 h^{\mu} - 2 \p^{\mu}\p^{\nu} h_{\mu\nu\alpha}\Box h^{\alpha} \right) \nn \\
&&-\,\frac{(9\, a + 5\, b)}4 \p \cdot h \Box \p \cdot h + (3\, a + b)\Big[ (\p^{\mu}\p^{\nu}h_{\mu\nu\rho})^2  +
\p^{\mu}\p^{\nu}\p^{\rho}h_{\mu\nu\rho} \p \cdot h \Big] \quad . \label{lab2} \eea

\no Since the term $h_{\mu\nu\rho} \Box^2 h^{\mu\nu\rho}$ is required in order to have a truly spin-3 content we
assume henceforth $a=1$ and rename the model as ${\cal L}_b \equiv {\cal L}_{(1,b)} $. It is invariant, for
arbitrary values of $b$, by the traceless reparametrizations (\ref{lbar}) determined by five gauge parameters
$\barl_{\mu\nu}$ which allow us to fix five gauge conditions.

Due to higher order time derivatives the analysis of the particle content of the free theory ${\cal L}_b $ is
nontrivial. Henceforth we follow the approach of \cite{deserprl}, see also  \cite{andringa,bhth}. We first fix a
gauge at action level, find a general solution of the gauge conditions in terms of helicity variables without
introducing time derivatives and plug it back in the action. Whenever we fix a gauge at action level we might
lose equations of motion which may not be recovered from the remaining equations of motion. According to the
recent references \cite{moto1,moto2} in order not to lose relevant equations of motion we are only allowed to
fix at action level the so called {\it complete}  gauge conditions. In our case this means that our five gauge
conditions must uniquely fix the five independent parameters $\barl_{\mu\nu}$ without any freedom for
integration constants. It can be shown that the five gauge conditions:

\be \p_{j} h_{jk\mu} = 0 \quad , \label{gc} \ee

\no where $ \, j,k = 1,2 \, , \,  \mu = 0,1,2$, are {\it complete}. Namely, requiring that the conditions
(\ref{gc}) are reached starting from arbitrary field configurations,

\be \p_j \left( h_{jk\mu} + \p_j\barl_{k\mu} + \p_k \barl_{j\mu} + \p_{\mu}\barl_{jk} \right) = 0 \quad, \label{geq}
\ee

\no we uniquely determine the five parameters\footnote{Recall that $\eta^{\mu\nu}\barl_{\mu\nu}=0$, therefore
$\barl_{00}=\barl_{jj} $ is not an independent variable.} $\barl_{k\mu}$ by repeatedly applying space
derivatives on (\ref{geq}). We obtain

\be \barl_{k\mu} = - \frac{1}{\nabla^2} \p_j h_{jk\mu} + \frac 1{\nabla^4} \p_{(k}\p_i\p_j h_{ij\mu )} -
\frac{\p_k\p_{\mu}}{3 \nabla^6} \left( \p_i\p_j\p_l h_{ijl} \right)  \quad . \label{lkm} \ee

\no The general solution to (\ref{gc}) is given in terms of five independent fields :

\bea h_{jkl} &=& \hat{\p}_j  \hat{\p}_k  \hat{\p}_l \psi \quad ; \quad h_{jk0} = \hat{\p}_j  \hat{\p}_k \phi \quad , \label{phi} \\
h_{00j} &=& \hat{\p}_j  \gamma + \p_j \Gamma \quad ; \quad h_{000} = \rho \quad , \label{rho} \eea

\no where $\hat{\p}_j = \epsilon_{jk}\p_k $ satisfies $\hat{\p}_i\hat{\p}_j = \nabla^2 \delta_{ij} - \p_i\p_j$
and $\hat{\p}_i\hat{\p}_i = \p_j\p_j = \nabla^2$. Back in (\ref{lab2}) with $a=1$ we have the decoupling of the couple $(\gamma,\psi)$ from the trio
$(\phi,\gamma,\rho)$,

\be {\cal L}_b = {\cal L}_{\gamma\psi} + {\cal L}_{\phi\Gamma\rho} \quad , \label{deco} \ee

\no where

\be {\cal L}_{\gamma\psi}= - (b+3) \gamma \nabla^6 \gamma + 2\, b\, \gamma \nabla^6 \Box \psi - (b+1) \psi
\nabla^6 \Box^2 \psi \quad , \label{lgp} \ee

\no and

\bea {\cal L}_{\phi\Gamma\rho} &=& - \frac{(b+1)}4\left( \trho \Box^2 \trho + 3 \, \trho \Box \nabla^2 \trho \right) - \frac 32 (b+1) \Gamma \nabla^2 \Box \dot{\trho} \nn\\
&-& 3(b+1) \trho \Box \nabla^4 \phi - (b+3) \trho \nabla^6 \phi - (b+3) \phi \Box \nabla^6 \phi \nn \\ &-&
3(b+3) \phi \nabla^8 \phi + 3 (b+3) \phi \nabla^6 \dot{\Gamma} - \frac 94 (b+1)\Gamma \Box \nabla^4 \Gamma \quad
, \label{lfgr} \eea

\no where $\trho = \rho - 3 \nabla^2 \phi$ and $\dot{f} = \p_{t} f$. If $b \ne -3$ we can write

\be {\cal L}_{\gamma\psi}= - (b+3) \tgamma \nabla^6 \tgamma - 4 \frac{(b+3/4)}{(b+3)} \psi \nabla^6 \Box^2 \psi
\quad , \label{lgp2} \ee

\no where $\tgamma \equiv \gamma - (b/(b+3))\Box \psi $ decouples from $\psi$. Due to the double massless pole
we have in general a massless ghost unless $b=-3/4$ where ${\cal L}_{\gamma\psi}$ has no particle content. On the other hand, if
$b=-3$, after the trivial shift $\gamma \to \bar{\gamma} + (2/3)\Box \, \psi$ in (\ref{lgp2}) we have

\be \left( {\cal L}_{\gamma\psi}\right)_{b=-3} = - 6\, \bar{\gamma} \nabla^6 \Box \psi = - \frac 32 \Big[
(\bar{\gamma} + \psi )\nabla^6 \Box (\bar{\gamma} + \psi ) - (\bar{\gamma} - \psi )\nabla^6 \Box (\bar{\gamma} -
\psi ) \Big] \quad . \label{bm3} \ee

\no Therefore, even if we change the overall sign of the Lagrangian, one of the two modes $\bar{\gamma} \pm
\psi $ is a ghost. Independently of what we find in ${\cal L}_{\phi\Gamma\rho}$ we  conclude that whenever
$b \ne - 3/4$ the Lagrangian ${\cal L}_b$ contains at least one ghost mode. This is in agreement with
the analysis made in \cite{dmmaster} for the case $b=-1$ which appears in the fourth order self-dual
model of that reference. From now on we focus on the only possible ghost free case: $b=-3/4$. After the redefinition

\be \Gamma = \tGamma - \dot{\trho}/(3\nabla^2) \quad, \label{gt} \ee

\no we get rid of the highest time derivatives present in (\ref{lfgr}). We perform another redefinition

\be \trho = \brho + 3 \nabla^2 \phi \quad, \label{brho} \ee

\no in order to cancel out all time derivatives of $\phi$. So we end up with

\bea {\cal L}_{b=-\frac 34} &=& - \frac{81}4 \phi \nabla^8 \phi - \frac 92 \phi\nabla^6\brho + \frac{27}4 \phi
\nabla^6\dot{\tGamma} - \frac 14 \brho \nabla^2 \Box\brho - \frac 9{16} \tGamma\nabla^4 \Box \tGamma \nn \\ &=&
- \frac{81}4 \bar{\phi} \nabla^8 \bar{\phi} - \frac 9{16} \bGamma \, \nabla^6 \, \bGamma  \quad , \label{34} \eea

\no where

\be \bar{\phi} = \phi - \frac{\dot{\tGamma}}{6\nabla^2} + \frac{\brho}{9\nabla^2} \quad ; \quad \bGamma =
\tGamma - \frac 2{3\nabla^2} \dot{\brho} \quad . \label{bfg} \ee

\no The equations of motion of (\ref{34}) lead to trivial solutions $\bar{\phi}=0=\bGamma $. Therefore, ${\cal
L}_{b=-\frac 34}$ has no particle content. Since we have made several changes of variables involving time
derivatives we should make sure that the two sets of fields $\Phi_K = (\phi,\Gamma,\rho)$ and $\bar{\Phi}_K =
(\bphi,\bGamma,\brho)$ are canonically equivalent. Notice that the diagonal form (\ref{34}) could have been
obtained at once from (\ref{lfgr}) at $b=-3/4$ via

\bea \phi &=& \bar{\phi} + \frac{\dot{\bGamma}}{9\nabla^2} - \frac{\Box}{9\nabla^4}\brho \quad , \label{c1} \\ \Gamma &=& - \dot{\bphi} + \left( 5 + \frac{\Box}{\nabla^2}\right) \frac{\bGamma}6 + \left(\frac{\Box + 3 \nabla^2}{9\nabla^2}\right) \dot{\brho} \quad , \label{c2} \\
\rho &=& 6 \nabla^2 \bar{\phi} + \dot{\bGamma} + \left(\frac{3\nabla^2 - 2\Box}{3 \nabla^4}\right) \brho \quad.
\label{c3} \eea

\no In matrix form we have $\Phi_J = \hat{M}_{JK}\bar{\Phi}_K$. The differential matrix  operator $\hat{M}_{JK}$
can be read off from (\ref{c1})-(\ref{c3}), it turns out remarkably that $\det (\hat{M})=1$. The reader can
check explicitly that $\hat{M}_{JK}\Phi_K=0 \to \Phi_k =0$. Therefore, $\Phi_J$ and $\bar{\Phi}_J$ are indeed
canonically equivalent. Moreover, the absence of $\brho$ in (\ref{34}) follows from a residual symmetry of the
gauge (\ref{gc})  at $b=-3/4$. At this specific point a transverse Weyl symmetry shows up. The residual symmetry
can be revealed by requiring invariance of the gauge (\ref{gc}) under $\delta h_{\mu\nu\rho} = \p_{(\mu}\barl_{\nu\rho )}
+ \eta_{(\mu\nu } \psi^T_{\rho )}$.

In summary, the fourth order Lagrangians  ${\cal L}_{(a,b)}$ contain a ghost mode for all values of $b$  except
$b=-3\, a/4$ where the model has  no propagating degree of freedom. Regarding the local symmetries, the
traceless reparametrization can be enlarged in only two cases. We can have longitudinal Weyl symmetry
$\delta h_{\mu\nu\rho} = \eta_{(\mu\nu}\p_{\rho )}\lambda $ at $b=-a$ or transverse Weyl symmetry
$\delta h_{\mu\nu\rho}= \eta_{(\mu\nu } \psi^T_{\rho )}$ at $b=-3\, a/4$. In the first case the six parameters
$(\barl_{\mu\nu},\lambda)$ can be combined into arbitrary reparametrizations governed by a traceful tensor
$\Lambda_{\mu\nu}$ which is however, not sufficient to make the theory ghost-free as we have seen.  In the second case we have
maximal symmetry with seven independent parameters $(\barl_{\mu\nu},\psi_{\mu}^T)$ and we end up with no particle
content.

Now we are ready to come back to investigate  the existence of a possible fourth order self-dual model connected
via some master action with our fifth order model (\ref{sd5}). The natural candidate for the fourth order term is
${\cal L}_{b=-3/4}$ for two reasons : it is the lower derivative term in ${\cal L}_{SD5}^{(3)}$ and it has no
particle content. However, if we built up a fourth order self-dual model only in terms of
a totally symmetric rank-3 tensor $h_{\mu\nu\rho}$ , i.e., ${\cal L}_{SD4}^{(3)} = {\cal
L}_{(a,b)}(h_{\mu\nu\rho}) + \cdots $, where the dots stand for lower derivative terms, then the equations of motion
$\delta S_{SD4}/\delta h_{\mu\nu\rho} = 0 $ must be of the form

\be E_{\mu}^{\s \beta}h_{\beta\nu\rho}^*(h) + E_{\nu}^{\s\beta}h_{\beta\rho\mu}^*(h) +
E_{\rho}^{\s\beta}h_{\beta\mu\nu}^*(h) + \cdots = 0 \quad, \label{eomg} \ee

\no where $h_{\beta\nu\rho}^*(h)$ is of third order in derivatives and stem entirely from ${\cal
L}_{(a,b)}(h_{\mu\nu\rho})$. Since we only have a totally symmetric field $h_{\mu\nu\rho}$ by hypothesis,
$h_{\beta\nu\rho}^*(h)$ must be also totally symmetric $h_{\beta\nu\rho}^*=h_{(\beta\nu\rho )}^*$. So if we apply
$\p^{\mu}\eta^{\nu\rho}$ on (\ref{eomg}), the resulting equation vanishes identically except for lower
derivative terms hidden in the dots. This means that the fourth order term ${\cal L}_{(a,b)}(h_{\mu\nu\rho})$
present in ${\cal L}_{SD4}^{(3)}$ must be invariant under longitudinal Weyl transformations\footnote{$\int d^3x
\, \lambda(\eta^{\mu\nu}\p^{\rho})\frac{\delta S_{SD4}}{\delta h_{\mu\nu\rho}} =0= - \frac{1}{3} \int d^3x
\frac{\delta S_{SD4}}{\delta h_{\mu\nu\rho}}\eta_{(\mu\nu}\p_{\rho )}\lambda = \int d^3x \frac{\delta
S_{SD4}}{\delta h_{\mu\nu\rho}} \delta_{\lambda} h_{\mu\nu\rho}.$}:

\be \delta_{\lambda} h_{\mu\nu\rho} = \eta_{(\mu\nu}\p_{\rho )}\lambda \quad . \label{lw} \ee

\no This is only possible if $b=-a$ which rules out the good candidate $b=-3\, a/4$. Therefore a possible fourth order self-dual
model made out of the $b=-3\,a/4$ term must have auxiliary fields besides $h_{\mu\nu\alpha}$ in order to avoid the above argument.
This is under investigation.

\section{Conclusion}

In $D=2+1$ dimensions, elementary particles  of helicity $ +1$ or $-1$ can be described either by the second order (in derivatives)  Maxwell-Chern-Simons (MCS) model of \cite{djt} or by the first order self-dual (SD) model of \cite{tpn}.   In the spin-2 case there are four, see \cite{ak1,desermc,djt} and \cite{sd4,andringa}, self-dual models ${\cal L}_{SDj}^{(2)} , \, j=1,2,3,4$ describing $+ 2$ or $-2$ helicity particles via $j$-th order theories. For spin-3/2 we have three self-dual models  of first, second \cite{deserkay,deser3/2} and third \cite{msl} order. In all those cases the $(j+1)$-th order model can be obtained from the $j$-th order one via master action and also via a Noether gauge embedding (NGE) procedure. There seems to be a ``$2s$'' rule regarding the highest possible order for a spin-$s$ self-dual model without ghosts. There is however, a caveat in the spin-3 case.

%In both works \cite{nge,dmmaster} we have succeeded in obtaining  ${\cal L}_{SD2}^{(3)} , {\cal L}_{SD3}^{(3)} $ and ${\cal L}_{SD4}^{(3)}$ from the first order model ${\cal L}_{SD1}^{(3)}$ of \cite{ak1}. Notwithstanding there are obstacles in both NGE and master action approaches which prevent us from going beyond the fourth order model even though  a ghost free sixth order model does exist \cite{bhth}. Here we have partially filled this gap by constructing a ghost free fifth order self-dual model.%

In section 2 we have revisited the derivation of the sixth order spin-3 self-dual  model ${\cal L}_{SD6}^{(3)}$ \cite{bhth}. The model is made out of one sixth plus one fifth order term. One can first obtain the fifth order term based on Weyl and arbitrary reparametrization invariances then, the sixth order term is obtained via a symmetrized curl of the fifth order term.

Next, in order to go one step downward and derive ${\cal L}_{SD5}^{(3)}$ we have started with the fifth order term of ${\cal L}_{SD6}^{(3)}$ and searched for a convenient fourth order term. Since lower order requires less symmetry, we have obtained a fourth order term by requiring traceless reparametrization invariance (\ref{lbar}) instead of general reparametrizations. This is also motivated by the key role of traceless reparametrizatons in higher spin theories in $D=3+1$. In particular, this is the symmetry behind the massless limit of the massive spin-3 Singh-Hagen model \cite{sh}. This requirement leads to the family of fourth order terms ${\cal L}_{(a,b)}$ in  (\ref{lab}). Imposing that the fifth order term is the symmetrized curl of the fourth order term uniquely determines $b=-3\, a/4$ where a transverse Weyl symmetry shows up. Consequently the new model ${\cal L}_{SD5}^{(3)}$ is uniquely determined. 

In section 3 we have connected ${\cal L}_{SD5}^{(3)}$ with ${\cal L}_{SD6}^{(3)}$ via master action which guarantees that the particle content of ${\cal L}_{SD5}^{(3)}$ is the same one of ${\cal L}_{SD6}^{(3)}$, i.e., massive particles of helicity $+3$ or $-3$ without ghosts. In section 4 we have investigated the possibility of going another step downward ${\cal L}_{SD5}^{(3)} \to {\cal L}_{SD4}^{(3)} $. A detailed study of  ${\cal L}_{(a,b)}$ reveals that only the case $b=-3\, a/4$ has no particle content and could be a good candidate to be the highest order term of a possible fourth order self-dual model ${\cal L}_{SD4}^{(3)} $. However, we have argued that it can not be entirely formulated in terms of a totally symmetric field $h_{\mu\nu\rho}$, auxiliary fields are required. We conjecture that only the self-dual models of order $2s$ and $2s-1$ can be formulated
in terms of totally symmetric rank-$s$ fields $h_{\alpha_{1}\ldots\alpha_{s}}$ without auxiliary fields. This is under investigation.

In summary, although we can obtain ${\cal L}_{SD6}^{(3)}$ from the new
fifth order model found here ${\cal L}_{SD5}^{(3)}$ as well as
the models ${\cal L}_{SDj}^{(3)}$, $j=2,3,4$  can be obtained from ${\cal L}_{SD1}^{(3)}$ of \cite{ak1},  there is
no connection between those two sets of models. The key point is that the fourth order term of ${\cal L}_{SD4}^{(3)}$
of \cite{nge} has a nontrivial particle content, so it can not be used to produce an equivalent 5th order model via master action.
From the point of view of local symmetries, the fourth order model of \cite{nge}
 is invariant under traceless reparametrizations plus longitudinal 
 Weyl transformations $\delta h_{\mu\nu\rho} = \p_{(\mu}\bar{\Lambda}_{\nu\rho )} + \eta_{(\mu\nu}\p_{\rho )}\Phi  $ which is equivalent to full reparametrizations $\delta h_{\mu\nu\rho} = \p_{(\mu}\Lambda_{\nu\rho )}$ while the model ${\cal L}_{SD5}^{(3)}$ found here  is invariant under traceless reparametrizations and transverse Weyl transformations $\delta h_{\mu\nu\rho} = \p_{(\mu}\bar{\Lambda}_{\nu\rho )} + \eta_{(\mu\nu}\psi_{\rho )}^T $. So there is no way of connecting those theories via Noether gauge embedding. 
 
 We believe that if we could be able to go all the way downward
from ${\cal L}_{SD6}^{(3)}$ until a first order model we would not end up at the  model  of \cite{ak1}.
There is probably a more natural first order spin-3 self-dual model which would allow us to go back upward until $\mathcal{L}^{(3)}_{SD6}$.
Eventually we might be able to construct first order self-dual models for arbitrary spin-$s$ in a more systematic way and learn more about higher spin theories.

\section{Acknowledgements}

The work of D.D. is partially supported by CNPq  (grant 306380/2017-0). A.L.R.dos S. is supported by a CNPq-PDJ
(grant 150524/2018-8) while the work of R.R.L. dos S. has been supported by FAPESP grants 2016/09489-0 and
2017/23966-9. D.D. thanks E.L. Mendon\c ca for a discussion.

\end{document}